# Spin wave localization and guiding by magnon band structure engineering in yttrium iron garnet

Rouven Dreyer[1], Niklas Liebing[1], Eric R. J. Edwards[1,†], Andreas Müller[1], and Georg Woltersdorf[1]*

[1]Institute of Physics, Martin Luther University Halle-Wittenberg, 06120 Halle, Germany

[†]Current address: IBM T. J. Watson Research Center, Yorktown Heights, New York 10598, USA

*E-mail: georg.woltersdorf@physik.uni-halle.de

*Abstract* – In spintronics the propagation of spin-wave excitations in magnetically ordered materials can also be used to transport and process information. One of the most popular materials in this regard is the ferrimagnetic insulator yttrium-iron-garnet due its exceptionally small spin-wave damping parameter. While the small relaxation rate allows for large propagation length of magnetic excitations, it also leads to non-locality of the magnetic properties. By imaging spin waves their band structure is mapped. In doing so wave vector selection is shown to suppress dispersion effects to a large extent allowing for local measurements of spin relaxation. Moreover we demonstrate even higher control of magnon propagation by employing the wave vector selectivity near an avoided crossing of different spin-wave modes where the group velocity approaches zero. Here local engineering of the dispersion allows constructing magnonic waveguides and at the same time reveals the local relaxation properties.

In recent years, spin-wave propagation and its control have been an intensely studied topic[1, 2]. In parallel, it has been demonstrated that spin waves may be used to transport heat[3] and angular momentum[4]. In many of these experiments, yttrium-iron-garnet (YIG) has proven to be a valuable material. The insulating properties of YIG were used in YIG/metal hybrid structures to demonstrate a flurry of magnetoresistive and magneto thermal phenomena which are explained by the excitation or annihilation of spin waves in YIG[5-9]. At the same time, the exceptionally small Gilbert damping constant of only $\alpha = 5 \times 10^{-5}$ of YIG enables spin transport on the millimeter length scale[3, 10]. In most cases, the presence of magnon excitations in YIG can be probed on the nanoscale by the inverse spin Hall effect[6, 11, 12]. However, this approach is not sensitive to the properties of the spin wave that is converted into a signal, i.e., its wavelength and propagation direction. For uniform magnetization dynamics (with wave vectors close to zero) ferromagnetic resonance is a reliable technique[13]. In the past decade, this method has been transferred to the micro- and nanoscale using vector network analyzers[14-17]. Spin waves with wavelengths down to 200 nm magnetization dynamics are investigated in a spatially resolved manner using micro-focus Brillouin light scattering (μBLS)[18-22] and scanning time-resolved magneto-optic Kerr microscopy (TR-MOKE)[23-27]. Beyond the resolution limit of visible light time-resolved X-ray imaging techniques can be utilized[28]. The long spin-wave relaxation times in YIG complicate the analysis since extrinsic effects such as sample inhomogeneity, magnon-magnon scattering[29], or instrumental effects related to the excitation of multiple spin-wave modes or limited frequency resolution usually dominate the measured line width[30]. On the other hand, the large spin-wave propagation length allows for investigating coupling phenomena such as avoided crossings in the spin-wave dispersion in single YIG films[31, 32] or in bilayer structures[33-35] consisting of insulating YIG and an adjacent conducting ferromagnetic material. In such systems the strong coupling regime between different magnon modes is accessible[36] paving the way for a novel playground for coherent information processing based on magnons[37]. Due to the large propagation lengths in YIG it is difficult to confine spin waves. Usually confinement is achieved by physically patterning (e.g. dry etching) the YIG material. Unfortunately, this approach introduces defects and modifies the magneto static properties. Therefore, for a number of experiments it would be highly desirable to have a method at hand allowing to control the spin wave properties locally without the need for patterning the YIG structure.

In this letter, we study the properties of spin waves in thin YIG layers by phase-resolved magneto-optic imaging of coherently excited spin waves. In order to reach the required frequency resolution and sensitivity for experiments with YIG a modified version of the TR-MOKE method is introduced. Besides the direct measurement of the spin-wave dispersion and avoided crossings of different spin-wave modes we demonstrate that a truly local measurement of spin-wave relaxation properties becomes possible. In addition, by extracting group velocities and relaxation times of the excited spin waves near an avoided spin-wave mode crossing we obtain an independent estimate for the local Gilbert damping parameter. Finally, we demonstrate that the avoided crossing results in a magnon band gap. Here we engineer the spin-wave dispersion locally to construct a soft magnonic waveguide which allows to study spin-wave propagation inside the spin-wave band gap opened by an avoided crossing of different spin-wave modes.

In our experiments we perform phase resolved magneto-optical microscopy of the magnetization dynamics on a 200 nm thick YIG layer. The magnetization is excited coherently using the rf-field generated by a coplanar waveguide (CPW) patterned on top of the YIG layer (as shown in Fig. 1a). The wave vector spectrum of the excited spin waves is determined by the static in-plane magnetic field as well as the frequency and spatial distribution of the excitation field. Due to the symmetry of the CPW the excitation field only has wave vector components perpendicular to the propagation direction along the waveguide. We distinguish between the Damon-Eshbach (DE) and backward volume (BV) configurations as limiting cases of their in-plane dispersion [31] c.f. supplemental Fig. S1. As light source a femtosecond laser operating at 520 nm with a repetition rate of $f_{\mathrm{rep}} = 80$ MHz is used to sample the magnetization dynamics via the polar magneto-optical Kerr effect (MOKE), as shown in supplementary Fig. S2. Due to the small spin relaxation, studying spin waves in YIG requires a method with a frequency resolution on the order of 1 MHz. To meet this requirement, we introduce a measurement scheme which we term super-Nyquist sampling MOKE (SNS-MOKE). The MOKE effect allows for mixing of the excitation frequency $f_{\mathrm{rf}}$ and the $n^{\mathrm{th}}$ harmonic of the pulse-repetition frequency yielding an intermediate frequency $f_{\mathrm{rf}} = n \cdot f_{\mathrm{rep}} + \varepsilon$. By demodulating the Kerr signal at frequency $\varepsilon$ directly yields real and imaginary components of the magnetic rf-susceptibility and in doing so provides a phase resolved measurements of the spin precession (see Supplemental Materials for details). Clearly, the advantage of the SNS-MOKE technique is that it allows for tuning of the rf-frequency in arbitrary steps (i.e. much smaller than the laser repetition rate of 80 MHz).

Typical data of the spatially resolved, complex susceptibility are shown in Fig. 1a) and recorded in DE and BV configurations, respectively. Here the external magnetic field is fixed at 142 mT and the rf-frequency is varied from image to image. The excited wave vector is determined by the maximum of the product of the **k**-dependent rf-magnetic field and rf-magnetic susceptibility $h(\mathbf{k}) \times \chi(\omega, \mathbf{k})$[30] and can be determined by counting the number of maxima $n$ observed over distance $L$ and calculating the wave number as $|\mathbf{k}| = 2\pi n/L$. In the following we use the spin-wave wavelengths $\lambda = |\mathbf{k}|/2\pi$ determined from spatially resolved images (Fig. 1a) or line scans to map out the dispersion of BV and DE modes by extracting the wave vectors as a function of frequency for a fixed magnetic field as shown in Fig. 2. Using the SNS-MOKE method we are able to map out the spin-wave dispersion with a pronounced avoided crossing [28, 38-41] using a step size of only 2 MHz, as shown in the inset of Fig. 2.

The avoided crossing of the first order perpendicular standing spin-wave mode (PSSW) and the DE mode has a much larger frequency splitting than the line width of individual spin-wave modes involved indicating strong coupling. Careful analysis following the formalism introduced by Kalinikos and Slavin[31, 42] allows to determine the mode repulsion of DE and first order PSSW mode and the corresponding size of the frequency splitting (see supplementary Fig. S3). In particular, we obtain a coupling constant $g = f_{splitting}/(\Delta f_1 + \Delta f_2)$ of 220 at 4 GHz. Values larger than $g = 1$ are referred to as strong coupling[43] and allow for coherent exchange of information between the two modes. The solid lines in Fig. 2 show the calculated dispersion for DE, BV (red) and PSSW (black) modes. In Fig. 2 the left inset shows the observed spatial profile of one of the hybridized modes in the vicinity of the avoided crossing. Here the propagation of the spin wave is strongly suppressed in comparison to the maps shown in Fig 1a). We attribute this effect to the nearly flat dispersion and therefore small group velocity close to the anti-crossing. Specifically, the spin-wave group velocity is reduced from 200 m/s to values of about 10 m/s near the mode repulsion (indicated by the purple arrow in the right inset of Fig. 2).

The Gilbert damping parameter is usually determined from the line width by sweeping the magnetic field at a fixed frequency across the ferromagnetic resonance (FMR). In Fig. 3 we record the SNS-MOKE signal in two-dimensional plots as a function of in-plane magnetic field magnitude and in-plane orientation. As one can see e.g. by following the signal along the dotted line locally measured field sweeps are not suitable to determine the Gilbert damping (c.f. suppl. Fig. S1). For the evaluation of these spectra, it would be required to take the wave vector distribution of the excitation field, the dispersion, and spin-wave propagation effects properly into account[44]. One might overcome this problem by identifying the magnetization direction where the dispersion is nearly flat and the spin waves cannot propagate to the probe spot. A nearly flat dispersion is expected for an intermediate angle of the field orientation between BV and DE configurations where the different dipolar contributions compensate each other (red dotted line in Fig. 2). The actual angle where the dispersion becomes nearly flat is indicated by black circles in Fig. 3 and depends on the rf-frequency (Fig. S4). A flat dispersion results in the simultaneous excitation of spin waves of all excited wave vectors causing destructive interference of all spin-wave modes except for the uniform mode (k = 0). This behavior is indeed observed in the inset of Fig. 3a), where mostly a uniform precession of the magnetization occurs. Note that the excitation of the uniform mode in the gap of the waveguide results only from the local out-of-plane rf-field. At 4 GHz, a flat dispersion is expected for an angle between **k** and **M** of 55° as indicated in Fig. 2. In addition, because the dispersion is flat, the excited spin waves have a nearly vanishing group velocity $v_{\mathrm{g}} = \partial\omega/\partial\mathbf{k}$ and cannot propagate. As demonstrated in the inset of Fig. 3e) indeed a pronounced minimum of the resonance line width is observed for these conditions. By measuring the susceptibility at the angles of minimal dispersion, we extract Lorentzian resonance line shapes from the local SNS-MOKE spectra that can be easily interpreted in terms of their line width. Fig. 4a) shows the frequency-dependent line width determined from a series of such spectra obtained in the point of minimal dispersion. The Gilbert damping determined from the red data points in Fig. 4a) corresponds to a value of α = (5.41 ± 1.07) x $10^{-5}$ with a very small zero-frequency line width offset $\Delta\mathrm{H}_0$ of only 16 μT and hence three times smaller compared to conventional FMR measurements we performed on the same sample. The remaining zero-frequency line width offset is most likely caused by two-magnon scattering processes which are facilitated by a flat dispersion extending to large **k**-vectors.

An even better localization of the probed properties may be achieved by controlling the magnon band structure such that degenerate magnon modes can be avoided completely. In the following, we will look in detail at the avoided crossing between the DE mode and the first PSSW mode and in doing so study spin waves with tunable propagation length with the aim to localize them. The coupling between different magnon modes leads to avoided crossings in the spin-wave dispersion, as shown in the right inset of Fig. 2. The associated change in curvature offers a handle to control the magnon propagation length since it is given by the product of spin-wave lifetime and group velocity[30]: $\lambda_{\mathrm{prop}} = \tau v_{\mathrm{g}}$, where the lifetime[45] is related to the Gilbert damping parameter by:

$$\tau = \frac{2}{\alpha} \frac{1}{\gamma \mu_0 \, (M_0 + H)} \tag{1}$$

Obviously, in the case of a flat dispersion the probed signal has a truly local character and provides access to intrinsic local properties such as internal fields and the Gilbert damping parameter. In the region of mode repulsion, the excited spin-wave modes are localized close to the edges of the conductors of the CPW (left inset in Fig. 2). Using different frequencies and a nearly flat dispersion found in the vicinity of the avoided mode crossing (shown in Supplementary Fig. S3) we extract the propagation length $\lambda_{\mathrm{prop}}$ and the group velocity $v_g$ for these modes and calculate an average Gilbert damping parameter of α = (4.08 ± 0.97) x $10^{-5}$ for frequencies between 1 GHz and 8 GHz using the equation above. The narrow line widths for different frequencies (Fig. 4a) and the corresponding Gilbert damping parameters (Fig. 4b) calculated from the obtained data are a consequence of the local character of the damping measurement under nearly flat dispersion conditions.

In the following we will demonstrate that the magnon band gap at small wave vectors created by the avoided crossing allows to realize magnon guiding along a track defined by tiny local fields. One can set the external field and the excitation frequency such that no spin waves can be excited and propagate (grey shaded area in e.g. Fig. 5a). In order to locally guide spin waves along a predefined track it is required to locally shift this band gap. For this we define magnetic structures on top of the YIG layer in order to provide a local bias field in their gap, as shown in Fig. 5c) and d). The local in-plane stray field amounts to ~ 0.5 mT for a 10 nm thick Permalloy structures with 5 µm gap and induces a shift of the magnon dispersion of about 100 MHz (Fig. 5b)). Now one can select a frequency where the spin-wave propagation within the magnonic waveguide is still allowed while it is forbidden due to the magnon band gap in the surrounding magnetic material. This situation is demonstrated in Fig. 5d) for a frequency of 4.45 GHz and an applied external field of 79 mT. The spin-wave mode within the magnonic waveguide can propagate over several 10 µm while the propagation on the left and right side of the waveguide is evanescent. By determining the decay length of this specific spin-wave mode and the corresponding group velocity from the dispersion we find a Gilbert damping parameter of $\alpha = (6.4 \pm 1.3) \times 10^{-5}$ (and calculate a line width of 10.3 ± 2.1 µT) (both indicated by orange stars in Fig. 4).

Summary. We demonstrate that by selecting the orientation of the wave vector it is possible to avoid spin-wave dispersion to a large extend. Even more interestingly, strong modification of the dispersion in the vicinity of the avoided crossing allows to select arbitrarily low values of the spin-wave velocity and hence to address the Gilbert damping locally. We demonstrate that the spin-wave interaction in the vicinity of the mode crossing represents a case of strong coupling. Indeed the dipole gap is a well-known regime which can host spin-wave Solitons[40] but may also allow to study cooperative phenomena such as Rabi oscillations for spin-wave excitations as well in the future. Finally, we demonstrated that a forbidden magnon band can be created by the avoided crossing. By local modification of the dispersion using the stray field of magnetic microstructures we have designed a "soft" magnonic waveguide. This waveguide supports spin waves in the vicinity of the avoided crossing while the propagation in the surrounding material is forbidden. Due to the low Gilbert damping tiny local bias fields are sufficient to control the propagation properties. In the future, such magnetic fields may be generated by currents flowing in microfabricated wire structures placed on top of the YIG layer allowing for a dynamic control of magnon propagation.

## Acknowledgement

Financial support from the German research foundation (DFG) through collaborative research center (CRC) 227 and Priority Program SPP 1538 (Spin Caloric Transport) and the European Research Council (ERC) via starting grant no. 280048 (ECOMAGICS) is gratefully acknowledged.

**Figures**

**Fig. 1**

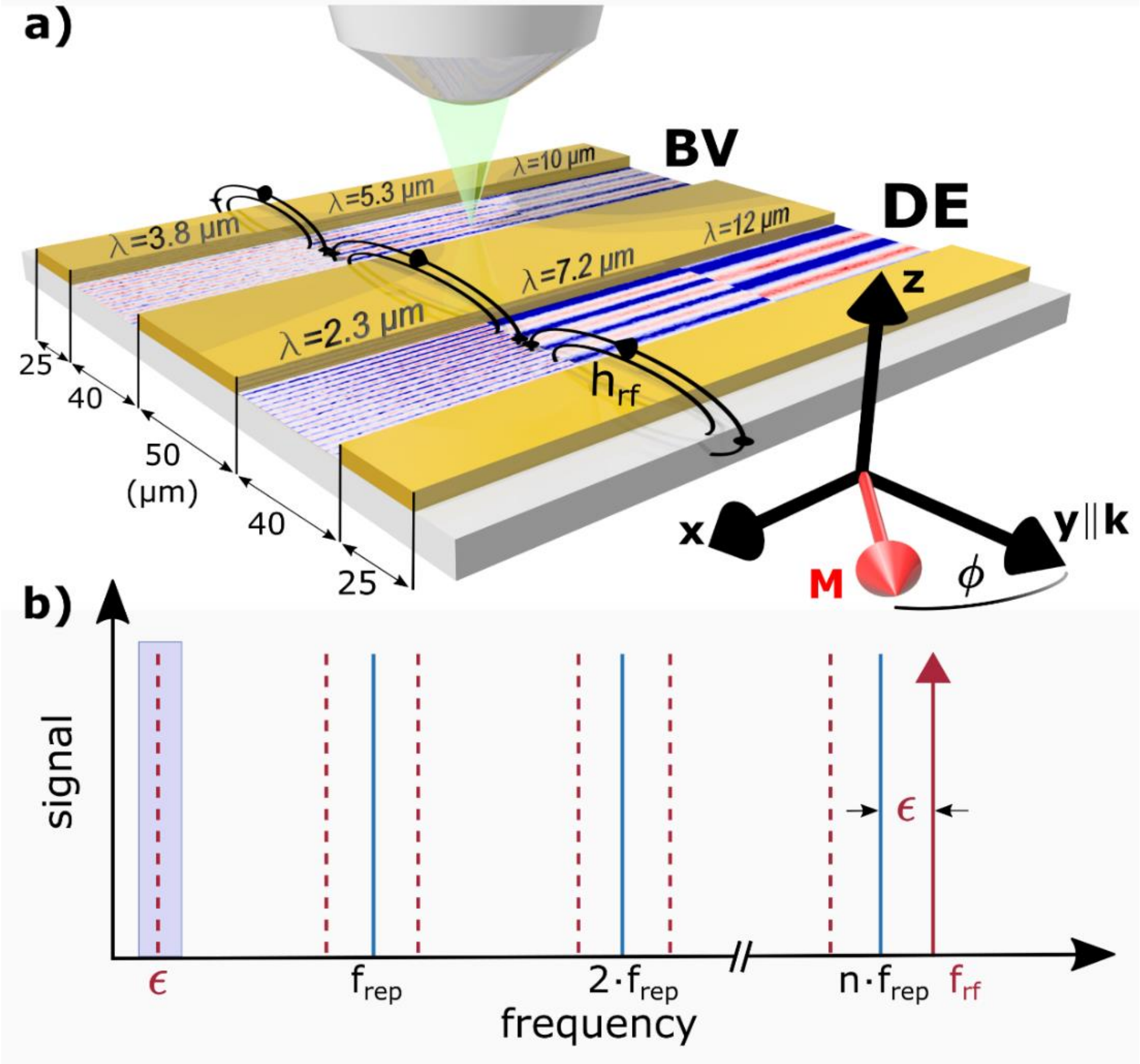

Fig. 1. a) Geometry of the experiments. Spatial resolved images of spin-waves in DE and BV configuration with their corresponding wavelength are shown in the gap. b) The frequency comb generated by a femtosecond laser is given by multiples of the laser repetition rate $f_{rep}$. The excitation frequency $f_{rf}$ aliases back to the Nyquist frequencies (red dotted lines). The lowest aliasing frequency corresponds to the difference frequency $\varepsilon$ between $f_{rf}$ and the nearest comb line of the frequency comb.

**Fig. 2**

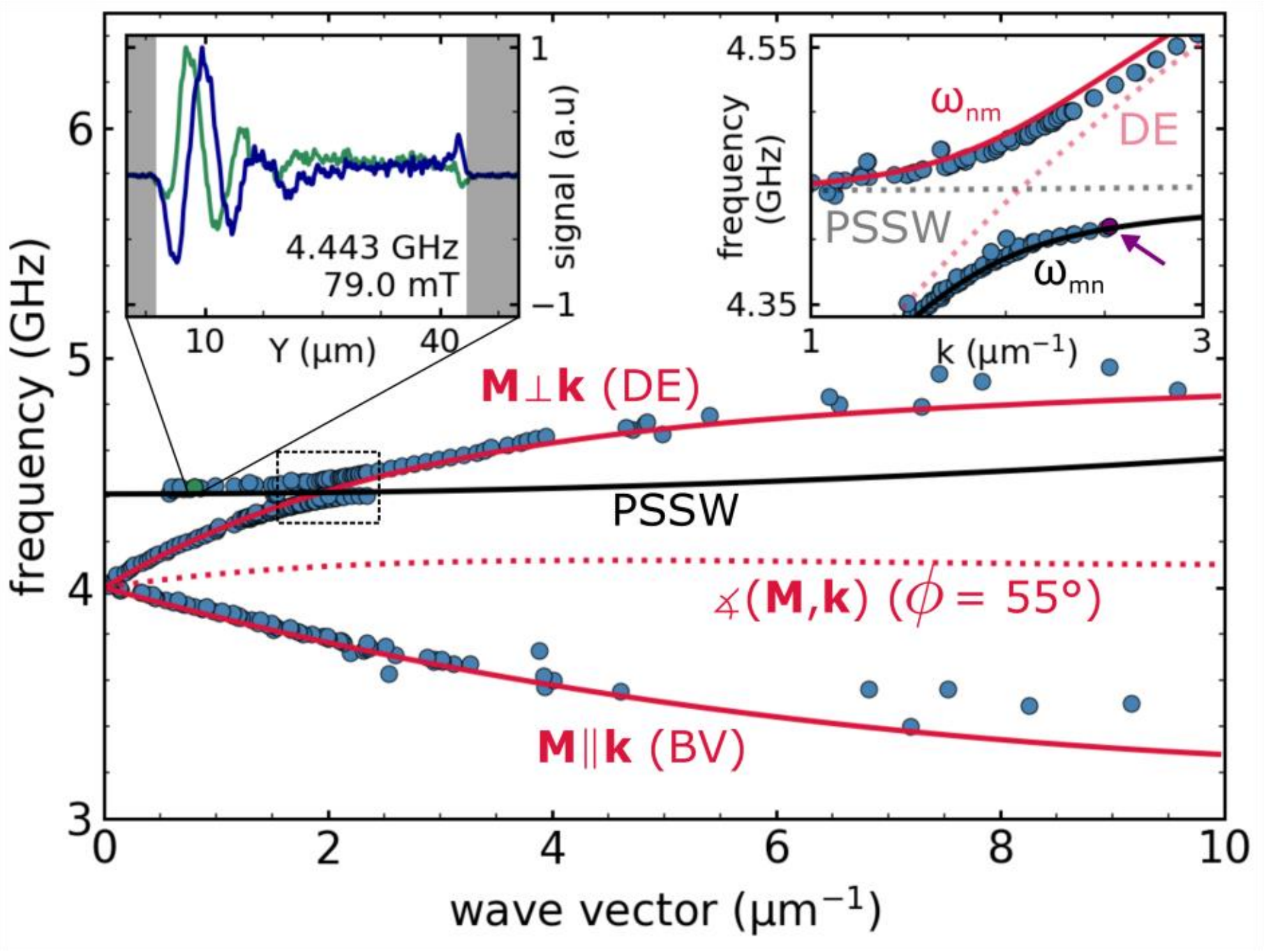

Fig. 2. Extracted spin-wave dispersion for a fixed external field of 79 mT. The dispersion curves were computed with the recipe by Harte[46] and Kalinikos[31, 47] The red solid lines indicate the DE and BV dispersion. The dotted red line shows an angular orientation of nearly flat dispersion. The black solid line depicts the dispersion branch for the first PSSW. The top left inset shows the spin-wave profile in the vicinity of the avoided mode crossing recorded in the gap of the CPW. The top right inset shows a magnified view of the avoided mode crossing with the calculated mode repulsion[31, 47]. All calculations were performed with the following parameters: thickness d = 192 nm, gyromagnetic ratio γ = 28.09 GHz/T[48], saturation magnetization $M_s$ = 139 kA/m[48], and exchange constant A = 3.6 pJ/m.

**Fig. 3**

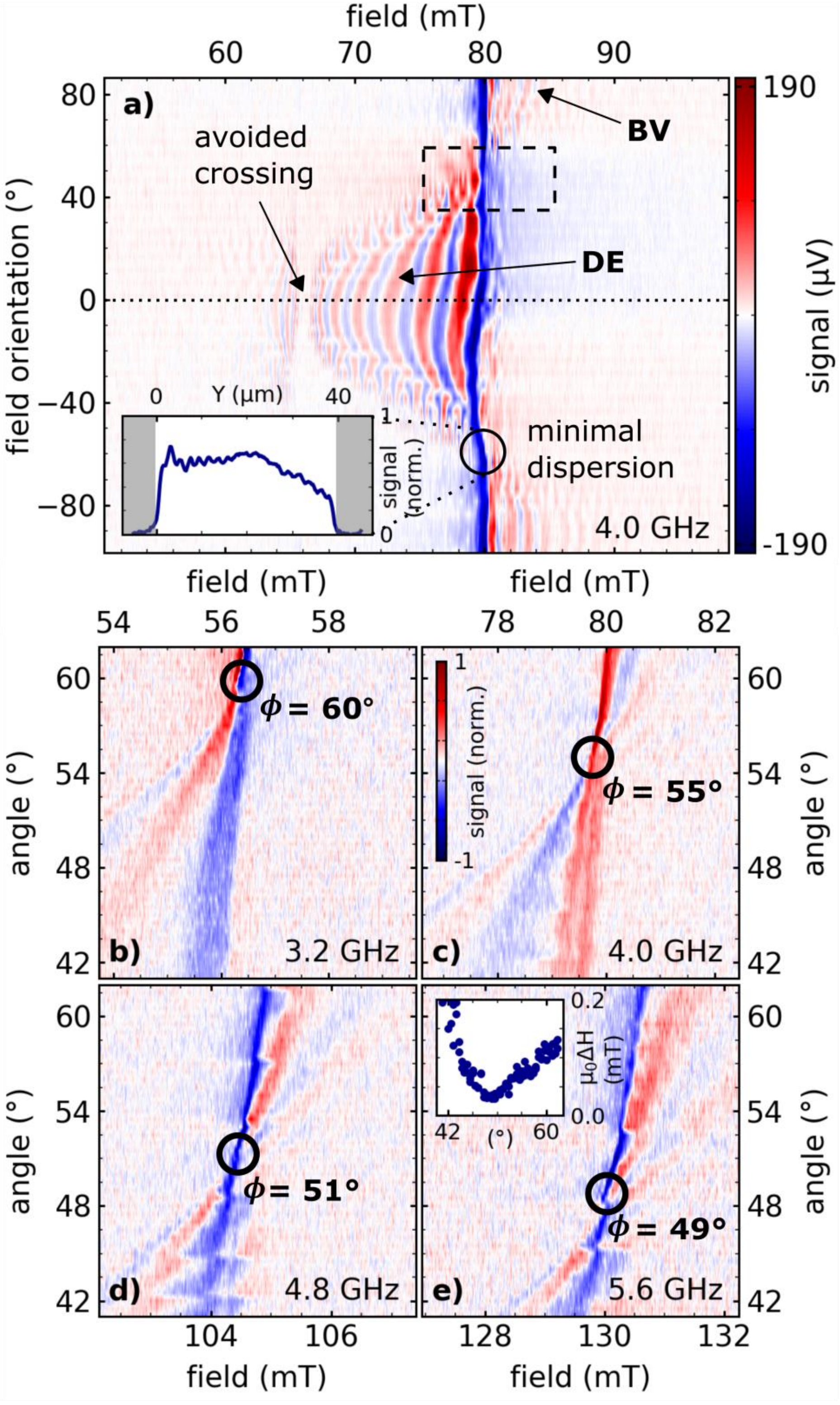

field (mT)
60
70
80
90
a)
avoided crossing
BV
DE
minimal dispersion
4.0 GHz
field orientation (°)
190
-190
signal (µV)
0
Y (µm)
40
signal (norm.)
field (mT)
φ = 60°
φ = 55°
φ = 51°
φ = 49°
angle (°)
signal (norm.)
b)
3.2 GHz
c)
4.0 GHz
d)
4.8 GHz
e)
5.6 GHz
µ0ΔH (mT)
(°)

Fig. 3. a) Angular dependence of the local field swept measurements of the rf-susceptibility measured with an excitation frequency of 4.0 GHz. A clear minimum of the dispersion around a magnetic field direction of $\Phi = 55\ °$ is visible. The inset shows the nearly uniform spatial distribution of the magnetic excitation across the gap recorded at the dispersion minimum. b)-e) Detailed measurement of the spin-wave dispersion minimum for frequencies between 3 and 6 GHz. The point of minimal dispersion shifts towards smaller angles with increasing frequency (see suppl. Fig. S3). The inset in panel e) depicts the extracted line width of the 5.6 GHz measurement as a function of in-plane field orientation clearly showing a minimum at nearly flat dispersion.

**Fig. 4**

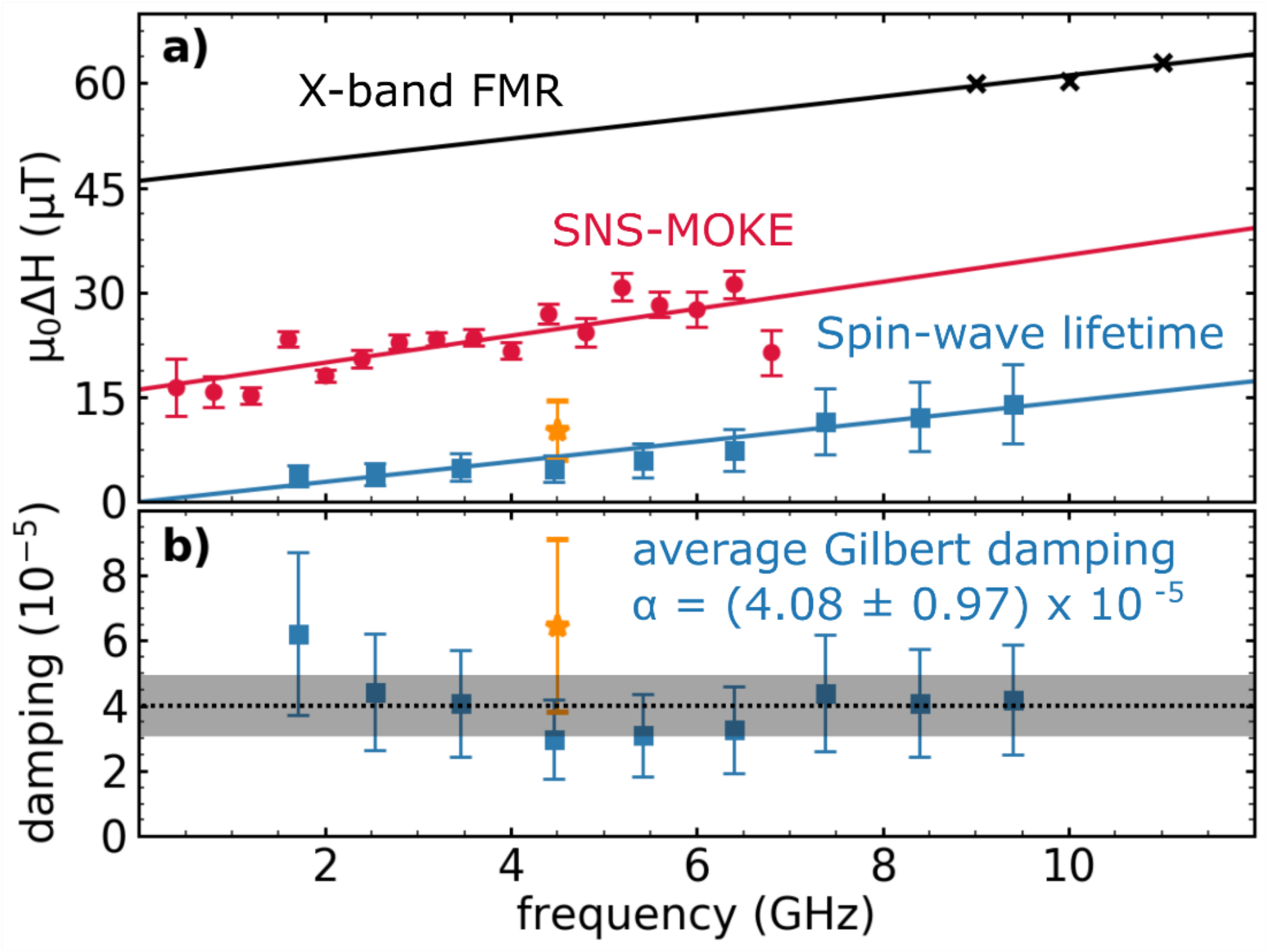

Fig. 4. a) Frequency dependence of the line width for FMR measurements (black cross) and at the dispersion minimum (red dots). The blue data points show the calculated line width in the vicinity of the avoided crossing Solid lines are fits to extract the Gilbert damping. b) Gilbert damping for localized spin-wave modes for different frequency. The dotted black line indicates the average damping parameter while the grey area is the standard deviation. The orange star marks the data point taken within the magnonic waveguide presented in Fig.5d).

**Fig. 5**

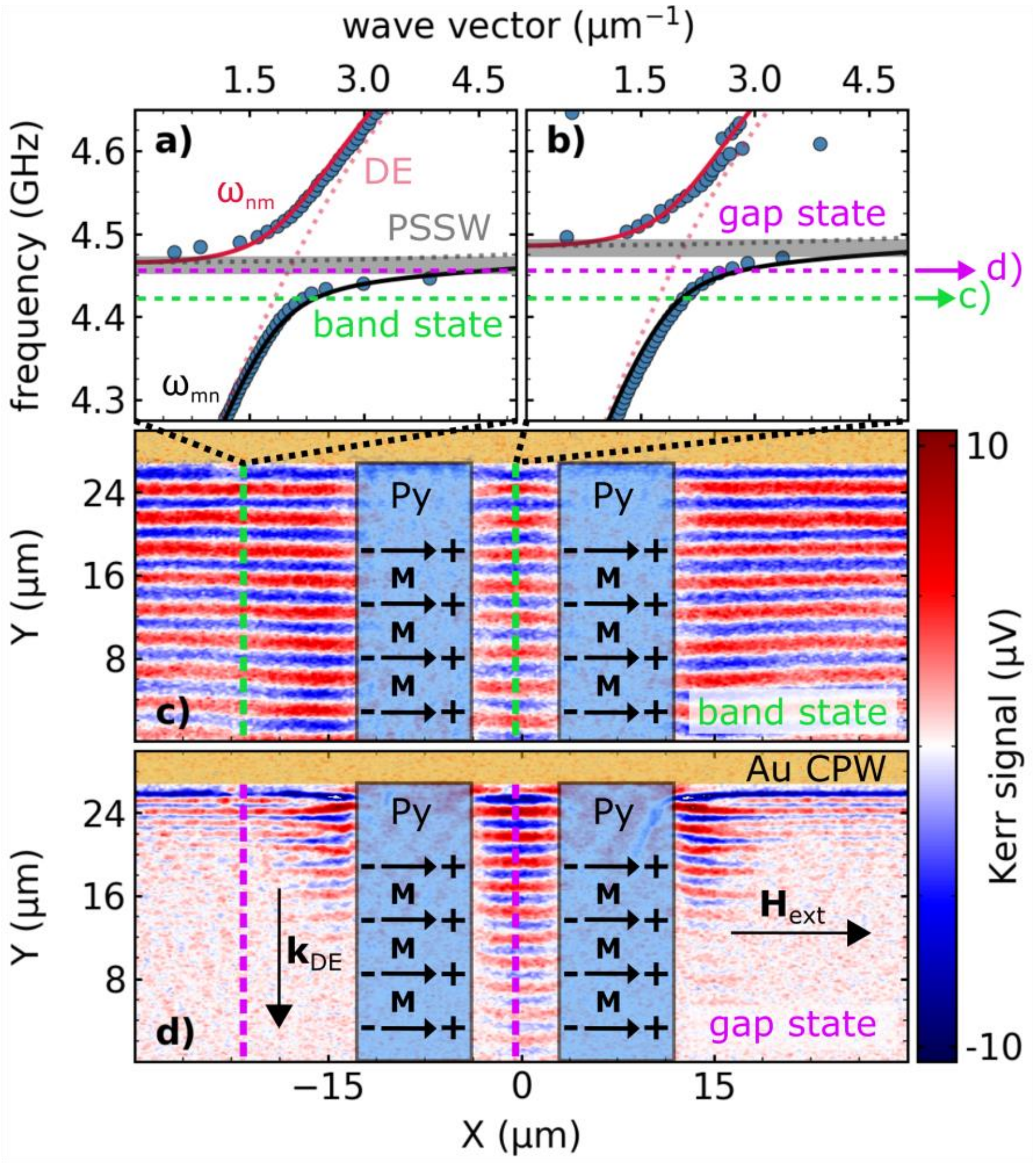

Fig. 5. The Py stripes next to ground line form a magnonic waveguide with a gap width of 5 µm. Within the Py stripes the magnetization aligns with the external field giving rise to stray fields at the edges of the stripe which enhances the external bias field within the magnonic waveguide. a) and b) show the magnonic band gap for an external field of 79 mT at positions next to and within the magnonic waveguide, respectively. In b) the dispersion is slightly shifted towards larger frequencies due to the stray field of the Py stripes indicated by the gray area. Here we find a regime where propagation of DE spin-wave modes is possible (gap state) within the waveguide but forbidden in the rest of the layer. In c) spin-wave propagation in both areas is allowed (band state). For a slightly increased frequency spin waves can only propagate within the magnonic waveguide as shown in d).

# Supplementary Material for

## Spin wave localization and guiding by magnon band structure engineering in yttrium iron garnet

Rouven Dreyer[1], Niklas Liebing[1], Eric R. J. Edwards[1,†], Andreas Müller[1], and Georg Woltersdorf[1*]

[1]Institute of Physics, Martin Luther University Halle-Wittenberg, 06120 Halle, Germany
[†]Current address: IBM T. J. Watson Research Center, Yorktown Heights, New York 10598, USA

*E-mail: georg.woltersdorf@physik.uni-halle.de

### Principle of super-Nyquist sampling MOKE

Typically, in TR-MOKE the rf-excitation and optical probing pulses are synchronized, sampling is stroboscopic such that $f_{\mathrm{rf}} = n \cdot f_{\mathrm{rep}}$ where $n$ is an integer, and the excitation is modulated (e.g., by microwave amplitude modulation) in order to allow for lock-in detection[1, 2]. This detection scheme limits the frequency resolution to multiples of the laser repetition rate (in our case 80 MHz) which rules out the observation of sub-MHz line widths and features within the spin-wave dispersion. Additionally, the inhomogeneous magnetization distribution of spin waves does not couple very effectively to the rf-field generated by the waveguide, hence a new approach compared to standard time-resolved magneto-optic microscopy (TR-MOKE)[3] with arbitrarily small frequency resolution is needed. Here, we detuned the rf-frequency from one of the laser comb lines such that:

$$f_{\mathrm{rf}} = n \cdot f_{\mathrm{rep}} + \varepsilon \tag{S1}$$

The dynamic response of the magnetization is down converted due to aliasing processes as shown in Fig. 1b) to the intermediate frequency $\varepsilon$. At non-zero $\varepsilon$ and with synchronized rf-excitation to the laser repetition rate, this down conversion occurs coherently, the phase information of the magnetization precession relative to the rf-excitation is preserved by lock-in demodulation at alias frequency $\varepsilon$. This process corresponds to an extreme case of undersampling or super-Nyquist sampling: spectrally narrow magnetization oscillations at the rf-excitation frequency in higher order Nyquist zones ($f_{\mathrm{rf}} \gg f_{\mathrm{rep}}$) can be reconstructed in an alias-free manner by exploiting the stability and the spectral width of the laser frequency comb. Thus SNS-MOKE does not require modulation of the excitation field or a delay mechanism and at the same time it overcomes the restriction of a fixed frequency grid given by integer multiples of $f_{\mathrm{rep}}$ for the excitation frequency. In addition, the SNS-MOKE approach simultaneously provides the real and imaginary components of the dynamic susceptibility whereas in standard TR-MOKE, the in-phase and out-of-phase parts need to be measured separately using a delay between the microwave signal and the laser pulses[1]. The simultaneously recorded complex signal components for the Damon-Eschbach and backward volume geometry are presented in Fig. S2 for a fixed excitation frequency of 6 GHz.

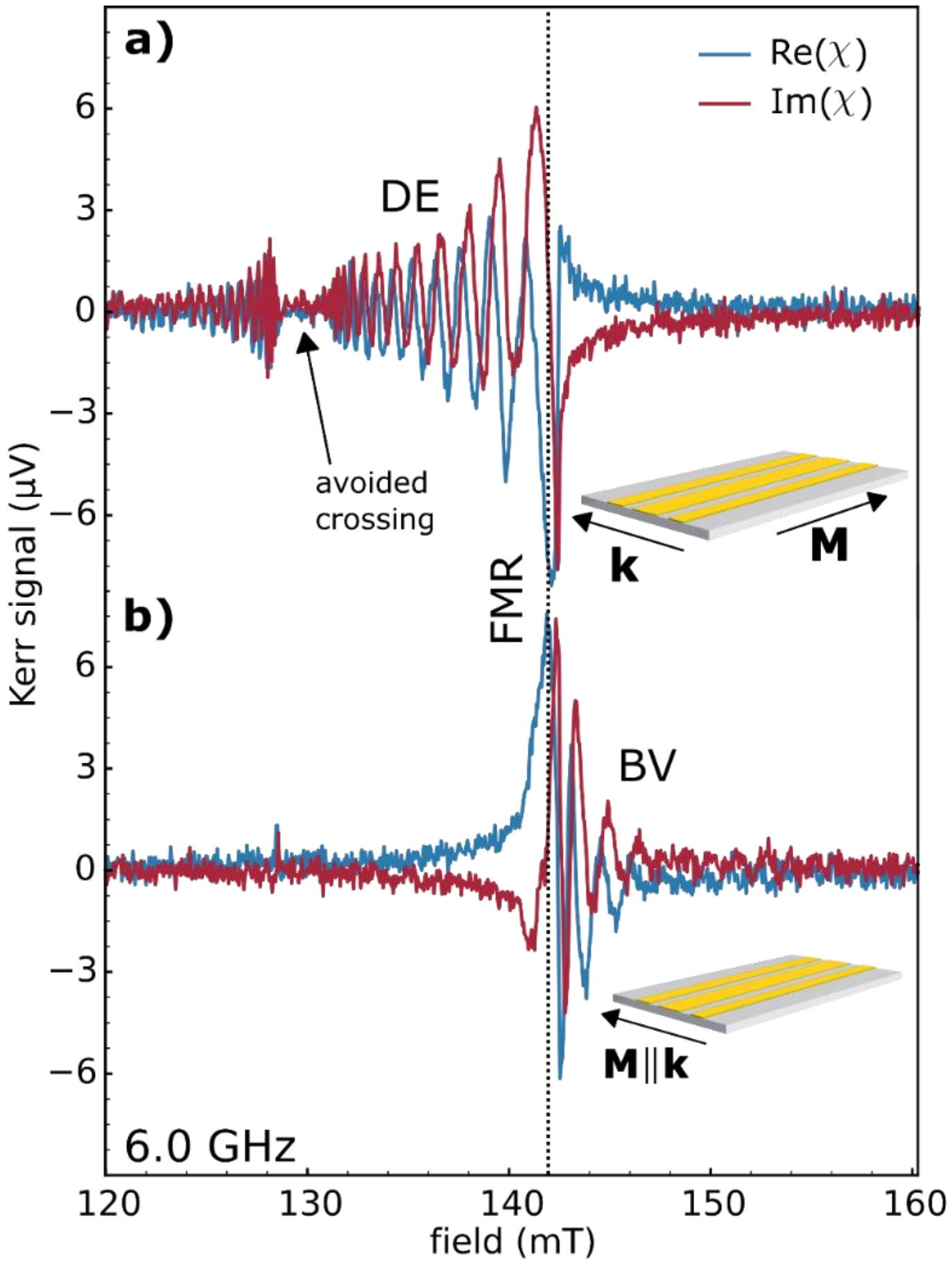

**Fig. S1. Spin-wave spectra in DE and BV configuration.** Local measurements of in-phase and out-of-phase components of the dynamic susceptibility a) with magnetization along the x-direction (DE configuration) and b) with magnetization along the y-direction (BV configuration). The microwave frequency was set to 6 GHz in both cases. The dotted black line indicates the resonance field of the uniform FMR mode at 142 mT. The suppression of the DE modes around 130 mT we link to an anti-crossing of the DE mode and the first order PSSW in the spin-wave dispersion[4, 5].

**Optical Setup**

In the presented experiments we utilize a femtosecond laser source operating at 520 nm to sample the magnetization dynamics in a spatially and phase resolved fashion (see Fig. S1). The laser spot on the sample is focused down to ~300 nm spot size with the aid of a high numerical aperture objective and the polarization state of the reflected light is analyzed by means of a balanced photodetector and Lock-In detection. The external magnetic field can be set with an electromagnet and the rf-excitation is provided by a micro-structured coplanar waveguide.

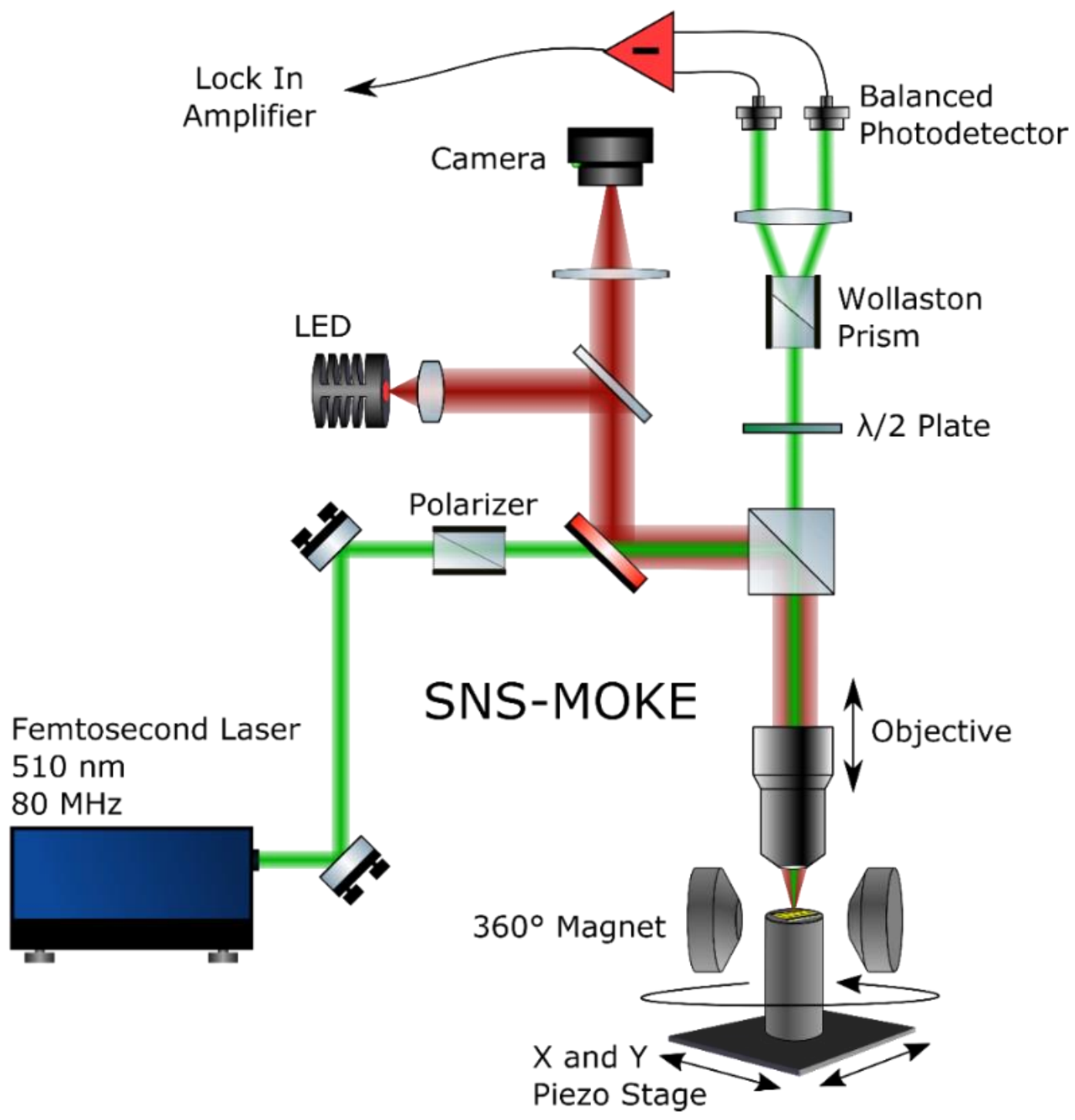

**Fig. S2. Experimental setup.** The SNS-MOKE setup consists of a femtosecond laser which is stabilized in its repetition frequency to an external master clock of 10 MHz. The polarization state of the optical pulses is maintained by using a Glan-Thomson polarizer before focusing the beam with a high numerical aperture objective lens onto the sample. By means of magneto-optical Kerr effect the polarization state of the reflected light is slightly changed due to the excited magnetization dynamics in the sample. This change of polarization is detected with a Lock-In amplifier at the corresponding alias frequency. The setup contains furthermore a rotatable electro magnet and a piezo positioning system for spatially resolved measurements. The position of the sample is observed with a camera system and actively stabilized there during the measurement.

**Simulation of the magnon dispersion.** For the calculations of the expected dipole gaps in the presented figures we utilized the formalism introduced by Kalinikos and Slavin[6, 7] for the case of an in-plane magnetized sample. The wave vector component along the film thickness of the DE mode $\kappa_{DE}$ is modelled by using the expression

$$[\kappa_{DE}^2 - d_1 d_2] \tan(\kappa_{DE} L) = \kappa_{DE}(d_1 + d_2) \quad \text{(S2)}$$

where $d_{1,2}$ are the pinning parameters of the two interfaces and $L$ is the film thickness. The wave number along the film thickness for the first order PSSW mode is given by $\kappa_{PSSW} = {}^{\pi}/_{L}$. In order completely model the mode repulsion it is necessary to first consider the expressions for the unperturbed dispersion branches for an in-plane magnetized film[7].

$$\omega_n^2 = (\Omega_n + \omega_M - \omega_M P_{nn})(\Omega_n + \omega_M P_{nn} sin^2\phi) \quad \text{(S3)}$$

Here, $\phi$ is the in-plane angle between magnetization and the wave vector of the excited spin wave, $\Omega_n = \omega_H + \omega_M l_{ex} k_n$ with $\omega_H = \gamma H$ and $\omega_M = \gamma M_S$ where $\gamma$ is the gyromagnetic ratio and $M_S$ saturation magnetization. Furthermore the exchange length $l_{ex}$ and the total wave number $k_n^2 = (k_{||}^2 + \kappa_n^2)$ (including the in-plane component $k_{||}$) are required. The expression for the dipolar contribution $P_{nn}$ is taken from the literature[6]. This formalism is used for the basic prediction of the pure DE, BV and PSSW dispersion branches shown in Fig. 2. The hybridized modes shown in Fig. 2 and Fig. 5 were determined with the following expression:

$$(\omega_n^2 - \omega_{nm}^2)(\omega_m^2 - \omega_{nm}^2) = \omega_M^2\ \Omega_n \Omega_m P_{nm}^2 \quad \text{(S4)}$$

The coupling and therefore the repulsion of the two modes is mediated by the interaction term $\omega_{nm}$ which results in the formation of an upper and a lower branch in the vicinity of the magnonic band gap. The magnitude of the frequency splitting is determined by the coupled dipolar contribution $P_{nm}$ including the in-plane and normal wave vector components of both modes and the pinning parameters[6]. To determine the parameters for the avoided crossing we utilized a Python based fitting routine which allows for simultaneous fitting of several data sets including avoided crossings obtained at different bias fields with the above shown formula for one set of fitting parameters. Doing so we obtained the following values for the exchange constant A = 3.6 pJ/m, saturation magnetization $M_s$ = 138 kA/m and thickness L = 192 nm for the YIG film for the different avoided crossings shown in Fig. S4. These values are in good agreement with the literature and the nominal thickness of the YIG film (200 nm). From the same fitting routine we extract the pinning parameters $d_1 = 1 \times 10^6$ m$^{-1}$ and $d_2 = 6 \times 10^6$ m$^{-1}$.

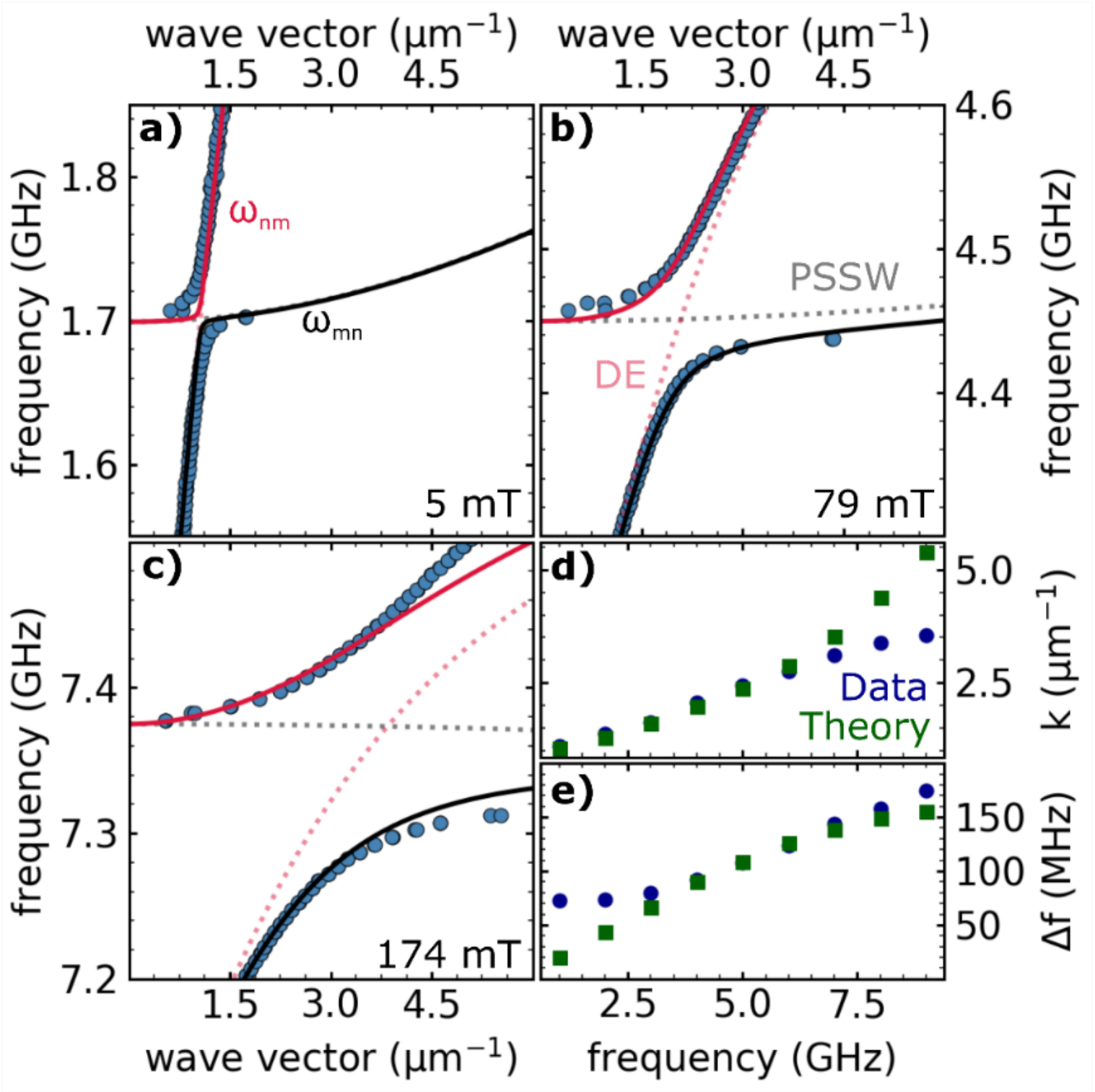

**Fig. S3. Avoided crossing in spin-wave dispersion.** In panel a)-c) the avoided crossing is shown for bias fields of 5 mT, 79 mT and 174 mT, respectively. The calculated branches $\omega_{nm}$ and $\omega_{mn}$ reproduce the measured dipole gap, while the dotted lines depict the calculated dispersion relation of the uncoupled DE (lightred) and the first order PSSW mode (grey). Panel d) shows the position of the avoided crossing for the experimental data (blue) und the calculated dipole gaps (green) as function of rf-frequency. In e) the size of the frequency splitting is compared with the theoretical expectations.

**Angular orientation for minimal dispersion.** To find the most suitable orientation for the linewidth analysis we performed angular resolved measurements over a wide frequency range and extracted the angular orientation of the minimal dispersion as function of frequency. As shown in Fig. S3 the orientation for a flat dispersion clearly shifts with frequency.

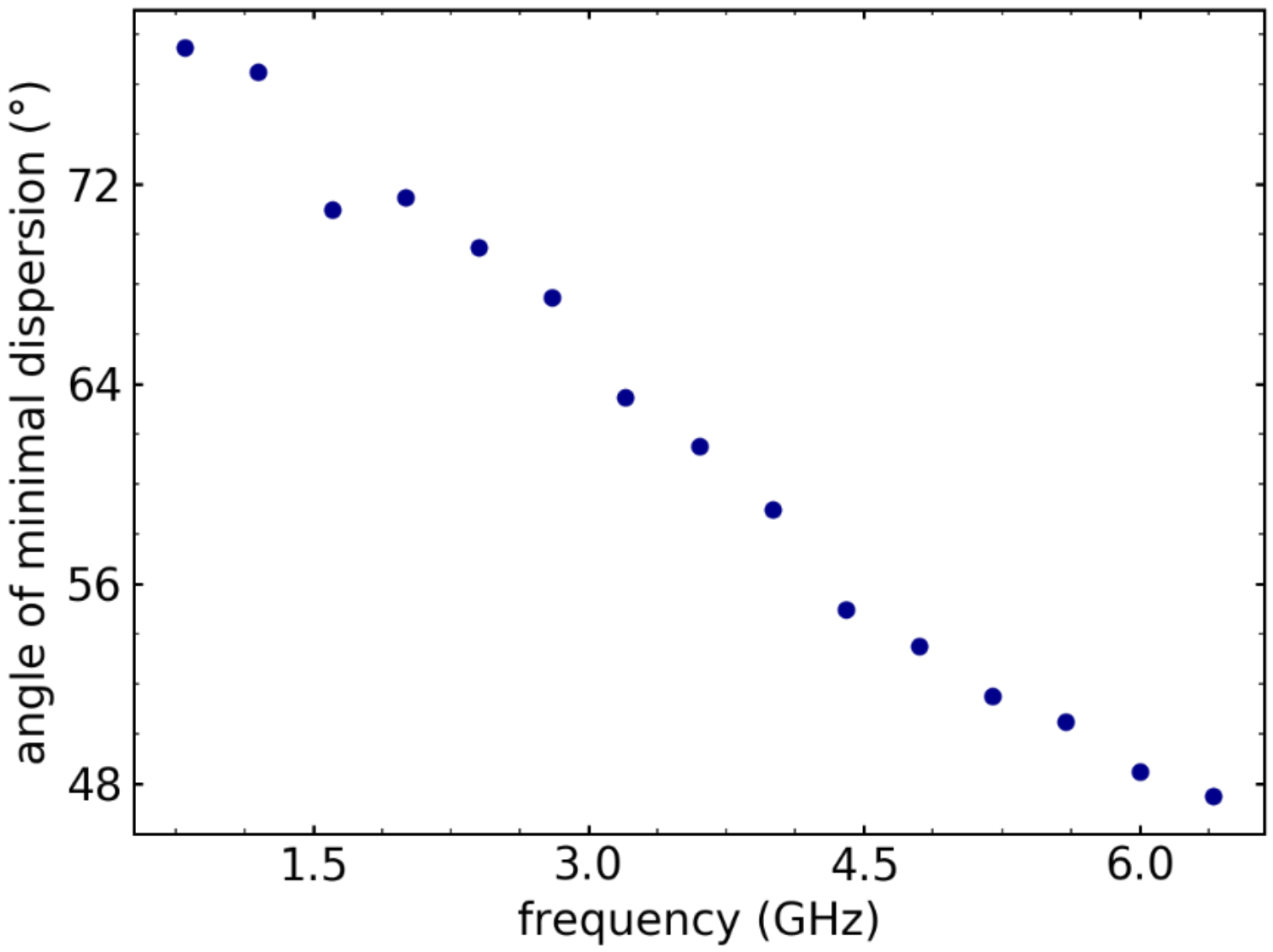

**Fig. S4. Frequency dependent change of minimal dispersion.** By performing angular resolved measurements as presented in Fig. 4 we determine the angular orientation with nearly flat dispersion in the magnonic band structure. The data points show the point of minimal dispersion for the frequency range of 0.8 – 6.4 GHz.

## Mode splitting under field reversal

Interestingly, one finds that the magnitude of the splitting depends on the direction of the magnetic bias field since DE modes have an amplitude profile with an exponential decay across the film thickness mostly localized at one of the interfaces (top or bottom) depending on the sign of the vector cross product of in-plane magnetic field and wave vector. For example, at 4 GHz reversing the magnetic field increases the splitting from 89 MHz to 102 MHz due to the different handiness of the DE modes (see Fig. S5).

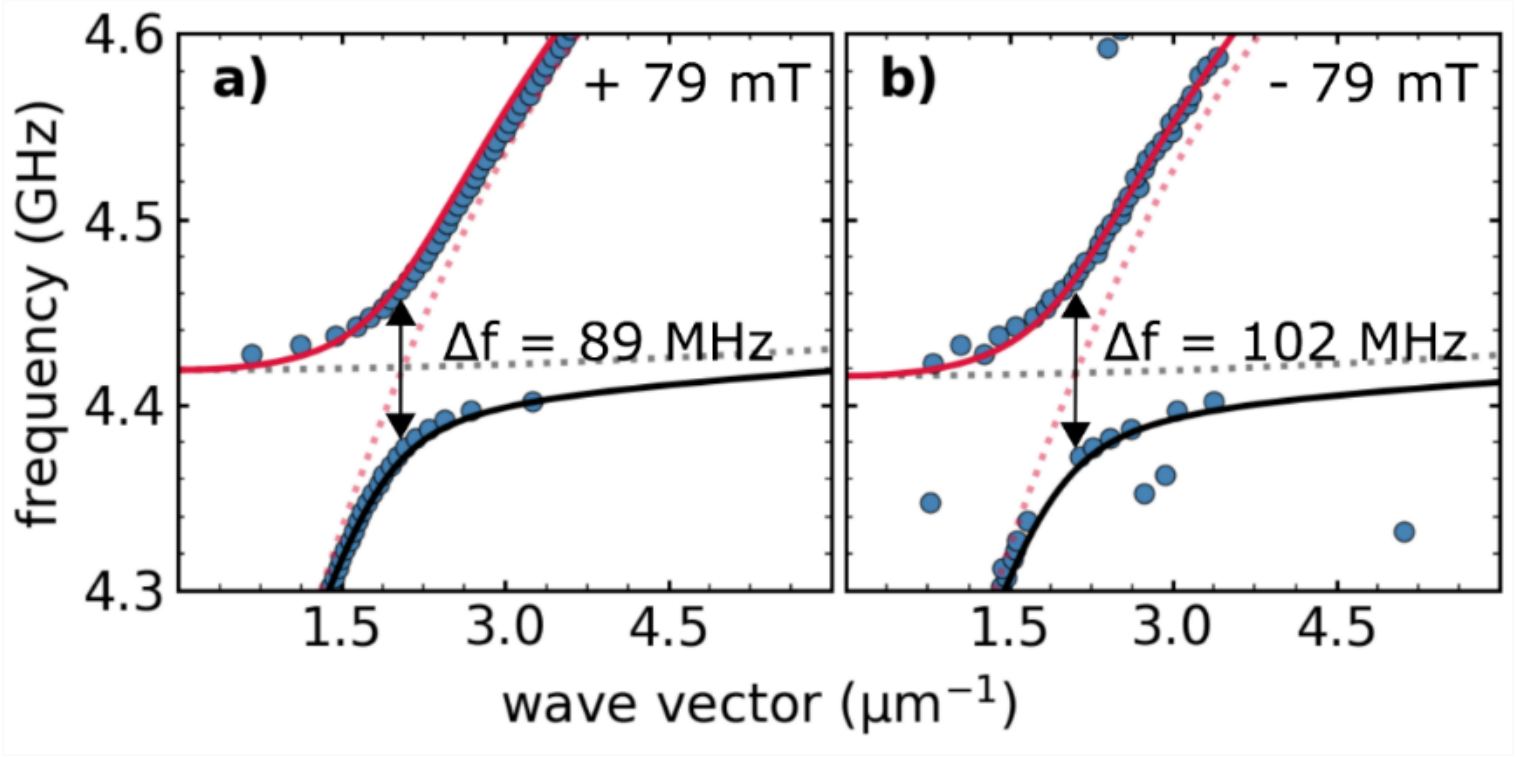

**Fig. S5. Change of frequency splitting under field reversal.** Measurement of the dipole gap at a fixed position close to the bottom groundline of the CPW with a) positive and b) negative field orientation. Changing the field direction from positive to negative leads to a more noisy set of data due to a reduced spin-wave amplitude in the line scans caused by the reversed handiness of DE spin-wave modes. At the same time the obtained frequency splitting is broadening by approximately 10 MHz.